\documentclass{article}

\usepackage{graphicx}
\usepackage{psfig}
\usepackage{epsfig}
\usepackage[round]{natbib}

\title{
Year-ahead prediction of US landfalling hurricane numbers
}

\author{Stephen Jewson}
\begin{document}

\author{Shree Khare\footnote{\emph{Correspondence address}: Email: \texttt{khare@ucar.edu}}\\and\\
Stephen Jewson\\}

\maketitle

\begin{abstract}
We present a simple method for the year-ahead prediction of the number of hurricanes making landfall in the US.
The method is based on averages of historical annual hurricane numbers,
and we perform a backtesting study
to find the length of averaging window that would have given the best predictions in the past.
\end{abstract}

\section{Introduction}

We are interested in developing more accurate methods for the prediction
of the number of hurricanes making landfall in the US during the North Atlantic hurricane season.
This question is of relevance to the many different groups of people who feel the effects of hurricanes,
such as homeowners, businesses, government and the insurance industry.
Accurate predictions, even if they are not \emph{highly} accurate,
can help such groups to make efficient use of resources for protecting against
the adverse effects of hurricanes.

A number of articles have presented methods for the prediction of the number of
Atlantic-basin hurricanes, such
as~\citet{blakeg04},
\citet{elsners93},
\citet{elsnerj04},
\citet{gray93},
\citet{gray94} and
\citet{klotzbachg03},
while others have
focussed on the prediction of the number of \emph{landfalling} hurricanes, such as~\citet{saundersl05}.
These studies all considered the question of how to predict the number of hurricanes
from a point in time either a few months before the start of, or during, the
hurricane season.
For instance in~\citet{gray93}
predictions were made from August 1,
in~\citet{gray94} predictions
were made from June 1, and in~\citet{saundersl05} predictions were made from August 1.
These kinds of predictions are usually known as \emph{seasonal} predictions, since they forecast
roughly a season in advance.
We, however, will consider predictions from a point in time
several months earlier than this,
at the end of the previous year's hurricane season. For some applications,
particularly in the insurance industry, predictions for this longer lead time
are more relevant than seasonal predictions.
Such \emph{year-ahead} predictions, as we call them (following~\citet{j51}),
as well as being useful in their own right, also serve as an important
baseline against which seasonal predictions should be compared.
Long term historical averages are not a reasonable
baseline for comparison since they are rather easy to beat even without a complex
seasonal prediction scheme, as we shall see below.

We take the idea of making year-ahead meteorological predictions from the
weather derivatives industry, where such
predictions are an essential part of the pricing of weather derivatives contracts
(see, for example~\citet{jewsonbz05}).
We will also take the forecasting method that we
present from that industry. The method is a very simple statistical
scheme that uses the observed time series of
annual historical hurricane numbers
to predict future values.

Meteorologists often use the word `climatology' to refer to the
unconditional statistics of the climate, and one could
refer to the predictions that we describe below as being methods for estimating the climatological mean
i.e. as methods for estimating the expected number of landfalling hurricanes before any dynamical forecasts
become available. The concept of time-independent climatology, however, although sometimes useful at a theoretical
or conceptual level, breaks down when one considers real meteorological observations.
No meteorological time series are stationary,
and every sensible estimate of likely future observations is always conditioned on past observations in some way.
We therefore prefer to think of all statements of likely future values of weather variables
as predictions of one sort or another, rather than as estimation of climatology.

In section~\ref{data} we describe the data on which this study is based,
in section~\ref{method} we describe the statistical method that we will use,
in section~\ref{results} we present our results and in
section~\ref{discussion} we discuss these results and mention some directions for future research.

\section{Data}
\label{data}

The data we use for this study is the HURDAT data-base\footnote{see~\texttt{http://www.aoml.noaa.gov/hrd/hurdat}},
that lists Atlantic hurricane occurences
from the late nineteenth century up to the present.
We will only use information about the number of hurricanes making landfall in the US (a variable
known as `\emph{xing}' in the database), and then only from 1900 onwards.
The data for landfalling hurricanes
for the period from 1900 to the present is generally believed to be reasonably accurate
(see, for example, the discussion on data quality in~\citet{elsnerk}).
In fact, our main test will only
use data from 1940 to 2004, while our sensitivity tests include the
(perhaps slightly less reliable) data from 1900 to 1939.

\section{Method}
\label{method}

The method we will use to predict lanfalling hurricane numbers is simple in the extreme:
we predict the number of landfalling hurricanes
in year $i$ as the average of the numbers of landfalling hurricanes per year over the $n$ previous years
(years $i-1,...,i-n$).
To calibrate the method we then use backtesting (also sometimes known as \emph{hindcasting})
to see which value of $n$ would have worked best in the past (over the period 1940-2004).
To make a forecast from our model we assume that the dynamics
of the time series have remained constant
(i.e. that the values of $n$ that worked well in the past will work well in the future)
and so we use the optimal value of $n$ to predict the number
of hurricanes for next year.
Our predictions are point predictions (i.e. single values): they can be taken either as a
prediction of the number of hurricanes or as an estimate of the \emph{expected} number of hurricanes.
In either case, the prediction is designed to minimise root mean square error (RMSE).

If we believed that the time series of hurricane numbers were stationary then we would
expect the optimal value of $n$ to be the
largest possible given the available data.
The use of such large values of $n$ appears to be quite common.
For instance, many of the authors who make seasonal forecasts have compared their
forecasts with forecasts made from such long term averages.
Also, the Florida Commission on Hurricane Loss Projection Methodology
(FCHLPM, a regulatory body that oversees the commercial risk modelling firms
that advise the insurance industry on hurricane risk) also
seems to suggest
that such a long term
average should be used to evaluate that risk.
\footnote{see \texttt{http://www.sbafla.com/}}.

\begin{figure}[!htb]
  \begin{center}
    \scalebox{0.8}{\includegraphics{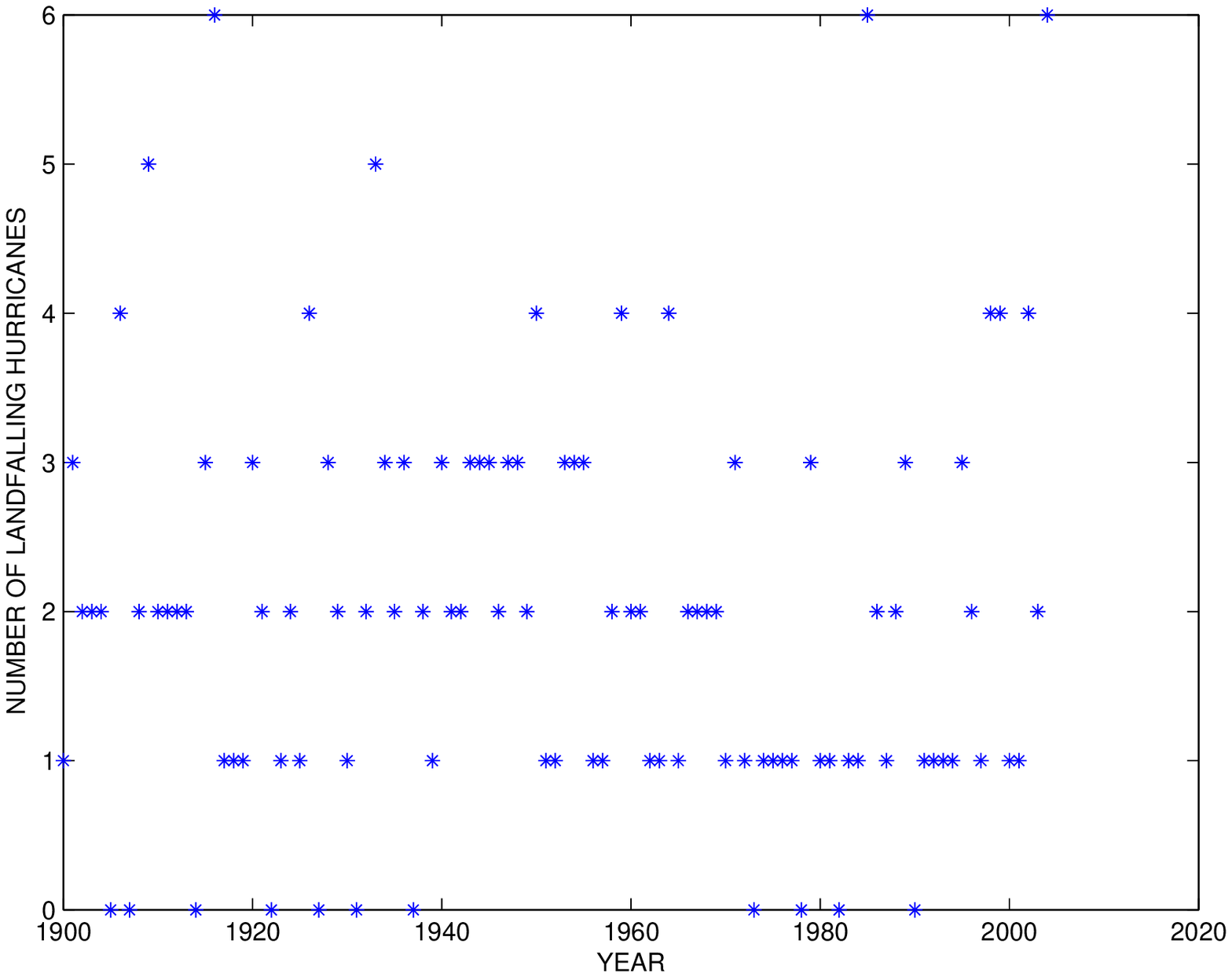}}
  \end{center}
  \caption{
The observed number of US landfalling hurricanes for each year since 1900.
          }
  \label{f1}
\end{figure}

However, there is abundant evidence that the landfalling hurricane number time series (as shown in figure~\ref{f1})
is \emph{not} stationary.
Firstly, it doesn't look stationary: there appear to be long time-scale fluctuations from decade to decade.
For example, the 1950's saw lots of landfalls, the 1970's and 1980's relatively few.
Secondly, there are good physical reasons to think that it isn't stationary.
One is that ENSO state has an impact
on hurricane numbers, and ENSO state fluctuates from year to year and decade to decade.
Another is that decadal time-scale fluctuations in
Atlantic sea surface temperatures probably have an impact on hurricane numbers
(see, for example, the discussion on this in~\citet{graysl97}).

It therefore makes sense to consider values of $n$ which are less than the longest possible,
and testing such lower values of $n$ is the goal of this study. \emph{A priori} we expect
that lower values will probably beat higher values of $n$ because of the non-stationarity
of the series.
However, the optimal value of $n$ would be very hard to determine
from physical principles
since it depends in a complex way
on the frequencies and amplitudes of the `signal' and the `noise'
in the hurricane number time series. However much physical understanding we have of the processes
governing hurricane formation it seems likely that it will only ever be possible to determine
the optimal $n$ empirically.

\section{Results}
\label{results}

\begin{figure}[!htb]
  \begin{center}
    \scalebox{0.8}{\includegraphics{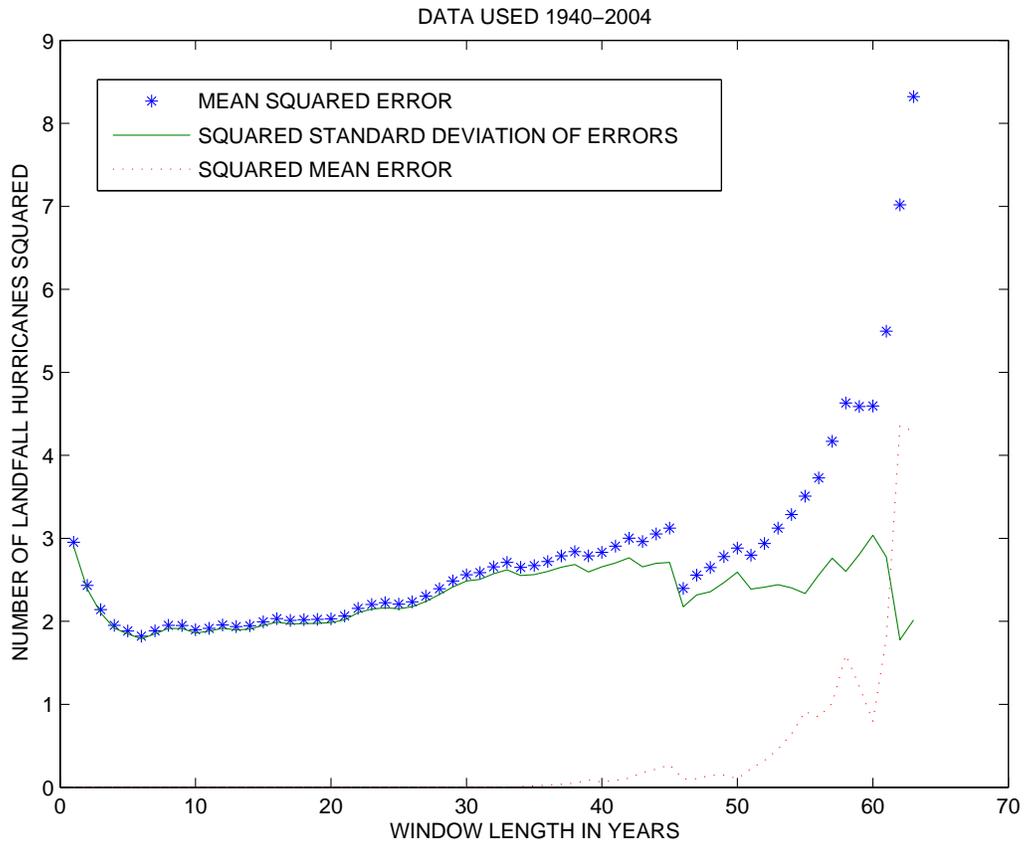}}
  \end{center}
  \caption{
The results from a backtesting study of the ability of averages of length $n$ years
to predict the time series of US landfalling hurricanes.
          }
  \label{f2}
\end{figure}

The results of our backtesting evaluation of different values of $n$ are shown in figure~\ref{f2}.
These results are based
on data from 1940 to 2004,
and show the MSE from all possible year-ahead hindcasts that one can make
using this data set, for each value of $n$.
The smaller the value of $n$ the more predictive tests are possible,
and hence the more terms in the sum that makes up the MSE.
Presumably the values of MSE for small $n$ are thus better estimated than those for large $n$.
The MSE values for small $n$ are also based on predictions over a wider range of years,
while results for large $n$ are necessarily based only on predictions of more recent data.

In addition to showing MSE (stars) we also show the break-down of MSE into the
standard deviation (SD) (solid line) and the bias (circles).

Considering the MSE results, we see that the lowest MSE values are given using a window length of 6 years.
Similarly low values of MSE are attained from window lengths either side of 6 years: values of $n$
from 3 years to 21 years all give MSE scores lower than 2.1.
The \emph{worst} results
come from the longest window length tested, of 64 years.
The breakdown of MSE into SD and bias shows that, except at the highest values of $n$,
the MSE is dominated by the SD.

We could, at this point, conclude that optimal forecasts of
future hurricane numbers should be made using a window of length 6 years, and
move on. However, there are three questions which deserve more investigation.
The first is whether it makes much difference to the forecasts generated what
length of averaging window is used, the second is whether
the minimum at 6 years
could have occurred
by chance even in a random stationary time series,
and the third,
given that the minimum of MSE is not particularly sharp,
is to investigate the sensitivity of the results to the number of years used in the analysis.

\begin{figure}[!htb]
  \begin{center}
    \scalebox{0.8}{\includegraphics{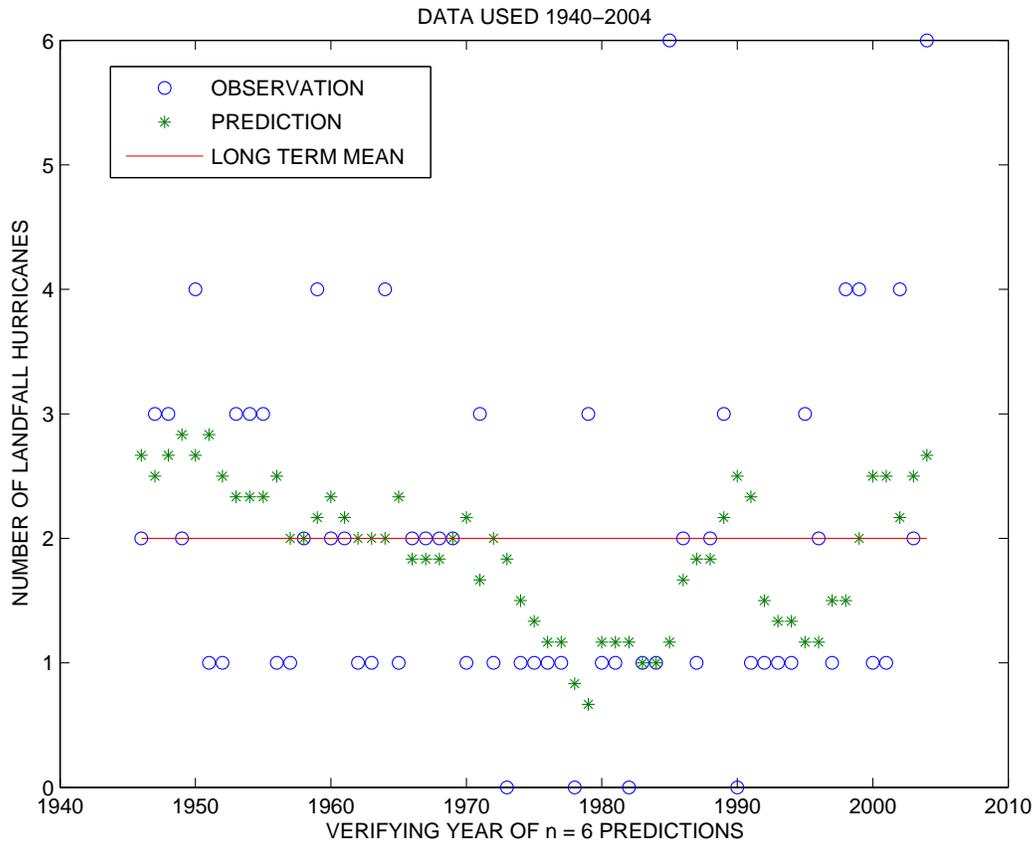}}
  \end{center}
  \caption{
Hindcasts of the number of landfalling hurricanes produced using an averaging window of length 6 years.
}
  \label{f6}
\end{figure}

With respect to the first of these questions, figure~\ref{f6} shows hindcasts produced
using an averaging window of length 6 years, along with actual hurricane numbers and the
long term average. We see that the hindcasts produced from the 6 year
averaging window fluctuate to a great degree, from values from near to 3 hurricanes per year
in the years around 1950,
to values below 1 hurricane per year in the late 1970s. There is a clear interdecadal
time-scale in these fluctuations.
Recent forecasts predict values just above 2, and are above
the long-term average. We conclude from this that it \emph{does} matter what method is used for year-ahead
prediction of hurricane numbers, since 6 year averaging and long-term averaging give
very different results.

\begin{figure}[!htb]
  \begin{center}
    \scalebox{0.8}{\includegraphics{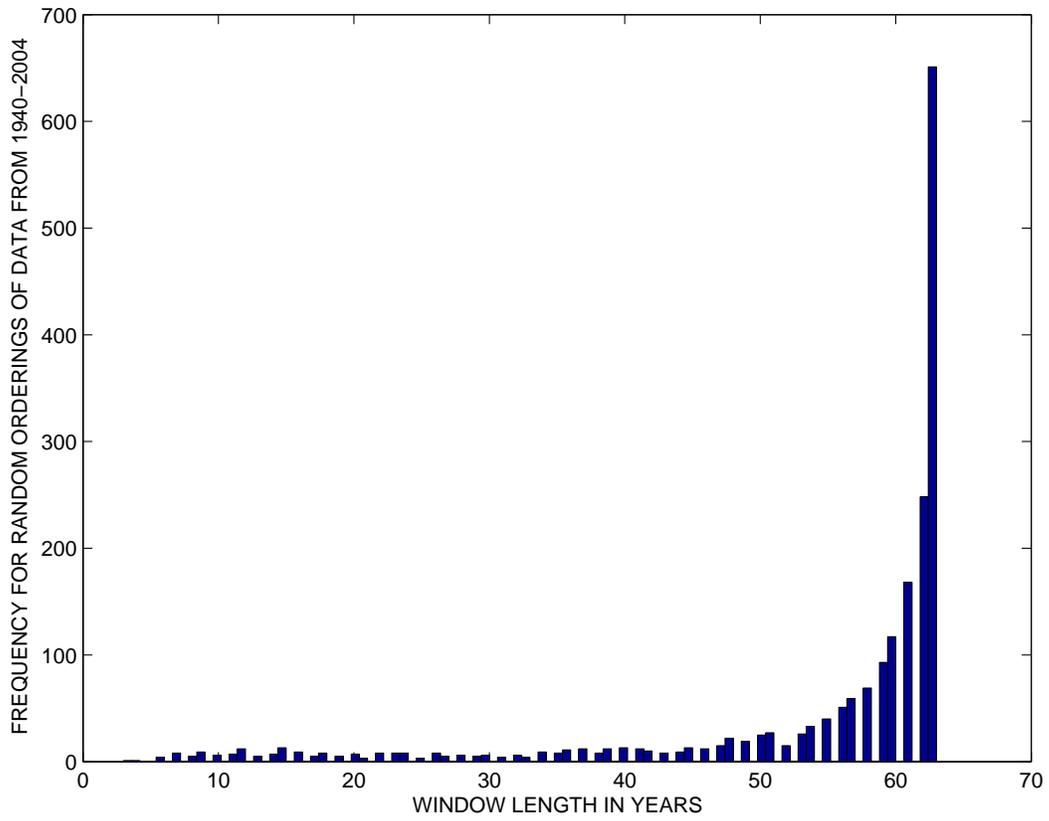}}
  \end{center}
  \caption{
The results from a statistical test of the minimum in figure~\ref{f2}.
Only 2 points fall below 6 years, giving a p-value of 0.1\%.}
  \label{f3}
\end{figure}

With respect to the second of these questions
(whether the minimum at 6 years could have occurred by chance)
we perform a simple statistical test as follows.
We randomly shuffle the hurricane number
time series 2,000 times, thus producing 2,000 new time series with the
same marginal distribution but different ordering. We then repeat
the backtesting analysis to find the optimal window length for each of these reshuffled time series.
This gives us 2,000 values of the optimal window length.
We compare the optimal window length for the real series (6 years)
against the distribution of optimal window lengths defined by these 2,000 values.
The distribution of the 2,000 optimal window lengths
is shown in figure~\ref{f3}. We see that this distribution has most of the mass at
much longer window lengths than 6 years. In fact, there are only
2 values below 6 years. We can thus calculate a p value of 2/2,000=0.1\% for this statistical test.
How should we interpret this result?
It seems to be strong statistical evidence that the observed time series is genuinely not
stationary: if it were then we would expect the optimal window length for the real series to be much longer than 6 years.

To address the third question, of the sensitivity of the value for the optimal window length versus
the number of years of data used, we perform two further tests.
First, we repeat the backtesting analysis using
data from 1900 to 2004.
As expected, given the short and noisy data series with which we are working,
the exact length of the optimal window length changes:
it is now 20 years rather than 6.
We also repeat the statistical test: this time 40 cases give values below 20 years.
This is less
statistically significant than previously (the new p-value is 40/2000=2\%),
but nevertheless confirms the general result that using
short rather than long windows
gives lower values of MSE.

\begin{figure}[!htb]
  \begin{center}
    \scalebox{0.8}{\includegraphics{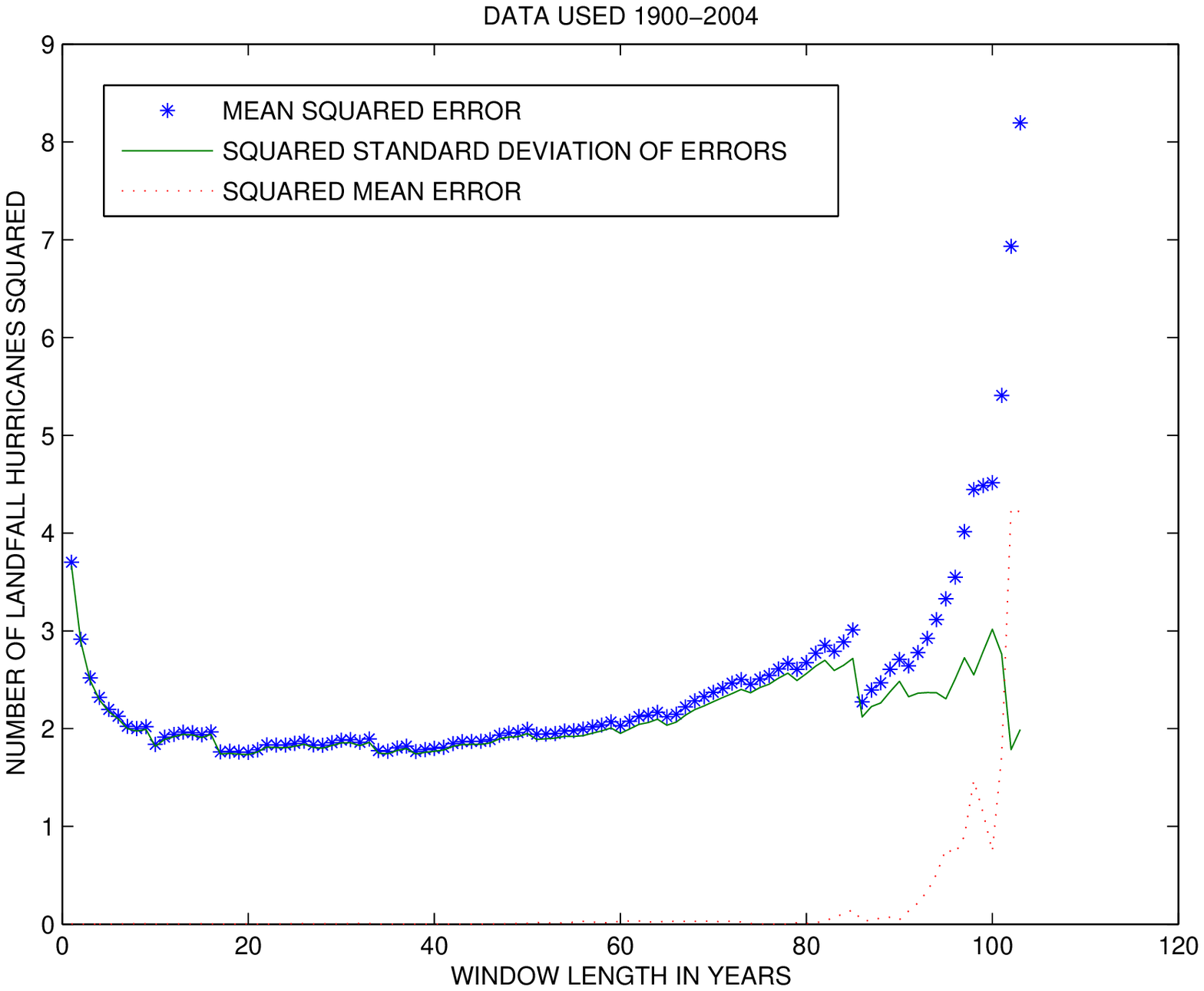}}
  \end{center}
  \caption{
As figure~\ref{f2} but for data from 1900 to 2004.
}
  \label{f4}
\end{figure}

\begin{figure}[!htb]
  \begin{center}
    \scalebox{0.8}{\includegraphics{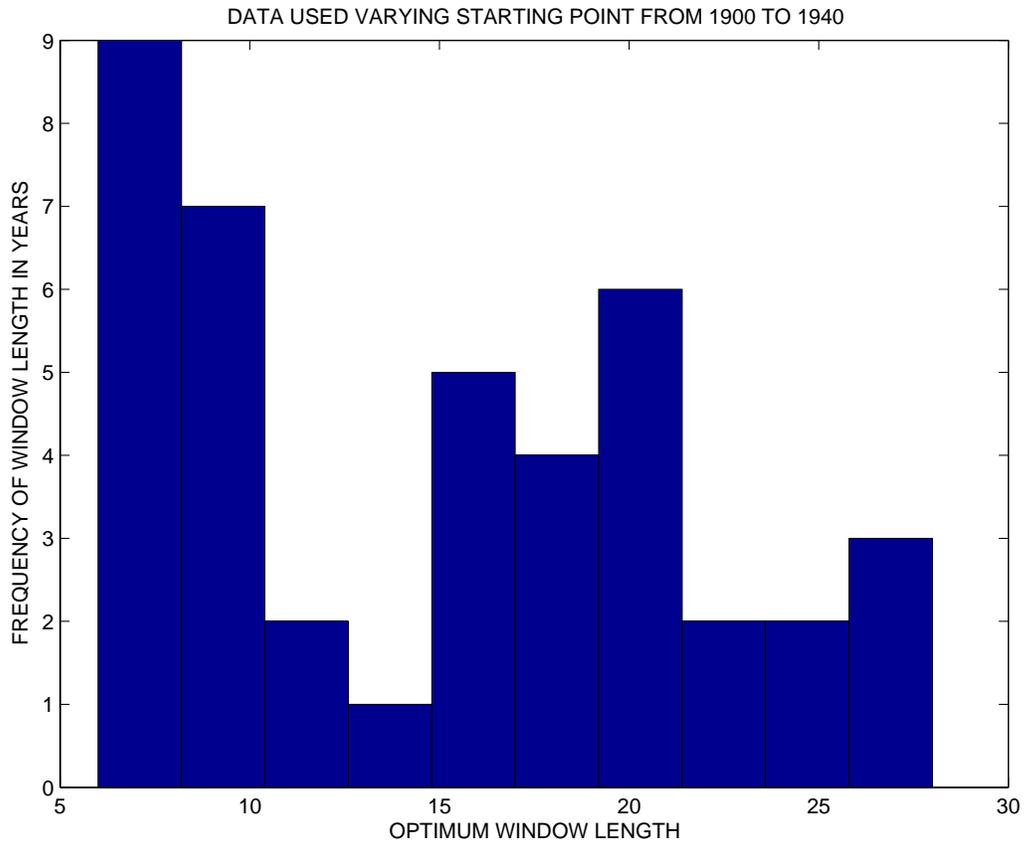}}
  \end{center}
  \caption{
The distribution of optimal window lengths from our backtesting comparison of methods for
prediction of the number
of landfalling hurricanes.
The shortest optimal window length is 6 years and the longest is 28 years.
}
  \label{f7}
\end{figure}

Secondly we repeat the backtesting analysis using all possible
starting points between (and including) 1900 and 1940. This gives us 41 backtesting
experiments. In each case we calculate the optimal window length, giving us 41 optimal
window lengths. The distribution of these 41 values are shown in figure~\ref{f7}.
We see a range of window lengths, from 6 to 28.
The result for the 1940-2004 data, giving a window of length 6,
is at one extreme end of the range. However,
in combination with the histograms in figures~\ref{f6} and~\ref{f4}
we again see clearly that
short window lengths (now in the range 6 to 28 years) are better than longer
window lengths.

\section{Discussion}
\label{discussion}

We have performed a simple investigation into how to forecast landfalling hurricane numbers
nearly a year in advance using time averages of the observed hurricane number time series.
In accordance with expectations, shorter averaging windows (of lengths 6 to 28 years)
perform better than longer averaging windows (of length more than 28 years).
This is presumably because the hurricane number time series is not stationary,
but shows long time-scale variations, and
short averaging windows capture information about the current phase of these
variations.

A statistical test based on reordering of the observations
shows that the short optimal window lengths that we derive
would be extremely unlikely to have occurred
by chance in a stationary random time series with the same marginal distribution
as the observations.
And sensitivity tests show that our results are not dependent on the exact period of historical data used
for the analysis.

Our method allows for the hurricane number time series to be non-stationary, but
assumes that the dynamics are constant (in the sense that we assume that predictions that have worked
well in the past will work well in the future). To the extent that the dynamics is
really changing (perhaps due to increases in the concentration of atmospheric
carbon dioxide) our results may not be terribly meaningful.
But then no other analysis of historical hurricane numbers would be meaningful either, and
we would be in a position where we could say almost nothing about the future.

This is our first foray into the question of how to predict hurricane numbers.
But there are a number of directions in which we intend to continue this research, including:

\begin{itemize}

    \item Testing schemes that model the interdecadal cycle in hurricane numbers using
    a trend model rather than just a flat-line

    \item Applying the methodology used above to model numbers of hurricanes in the whole North Atlantic basin,
    numbers of hurricanes in different parts of the basin, the proportion of hurricanes making
    landfall, weak and strong hurricanes, tropical storms in other basins, and so on.

    \item Extending the method to include a `detrending' element,
    in which we calculate optimal estimates of the long time scale fluctuations that are presumably the source
    of the year-ahead predictability that we have detected

    \item Applying the methodology to other important climate time series, such as the NAO.

\end{itemize}

In this study we have considered point (single value) predictions of the number of hurricanes.
However, for many applications, predictions of the whole distribution of the
possible number of landfalling hurricanes would be more useful.
To this end, we also intend to extend the methodology to predict this distribution.

Finally we note that one of the loose ends from this study is that we have only
pinned down the optimal window length to lie within a range from 6 to 28 years.
A method that merges predictions from different window lengths
(perhaps using likelihood-weighted model averaging) might be a reasonable way to
improve our simple prediction methodology, and would avoid having to make an
arbitrary choice of window length from within this range of values.

%

\section{Legal statement}

SJ was employed by RMS at the time that this article was written.

However, neither the research behind this article nor the writing
of this article were in the course of his employment, (where 'in
the course of their employment' is within the meaning of the
Copyright, Designs and Patents Act 1988, Section 11), nor were
they in the course of his normal duties, or in the course of
duties falling outside his normal duties but specifically assigned
to him (where 'in the course of his normal duties' and 'in the
course of duties falling outside his normal duties' are within the
meanings of the Patents Act 1977, Section 39). Furthermore the
article does not contain any proprietary information or trade
secrets of RMS. As a result, the authors are the owners of all the
intellectual property rights (including, but not limited to,
copyright, moral rights, design rights and rights to inventions)
associated with and arising from this article. The authors reserve
all these rights. No-one may reproduce, store or transmit, in any
form or by any means, any part of this article without the
authors' prior written permission. The moral rights of the authors
have been asserted.

The contents of this article reflect the authors' personal
opinions at the point in time at which this article was submitted
for publication. However, by the very nature of ongoing research,
they do not necessarily reflect the authors' current opinions. In
addition, they do not necessarily reflect the opinions of the
authors' employers.

\bibliography{shree2}

\newpage

\newpage

\newpage

\newpage

\newpage

\newpage
\begin{figure}[!htb]
  \begin{center}
    \scalebox{0.8}{\includegraphics{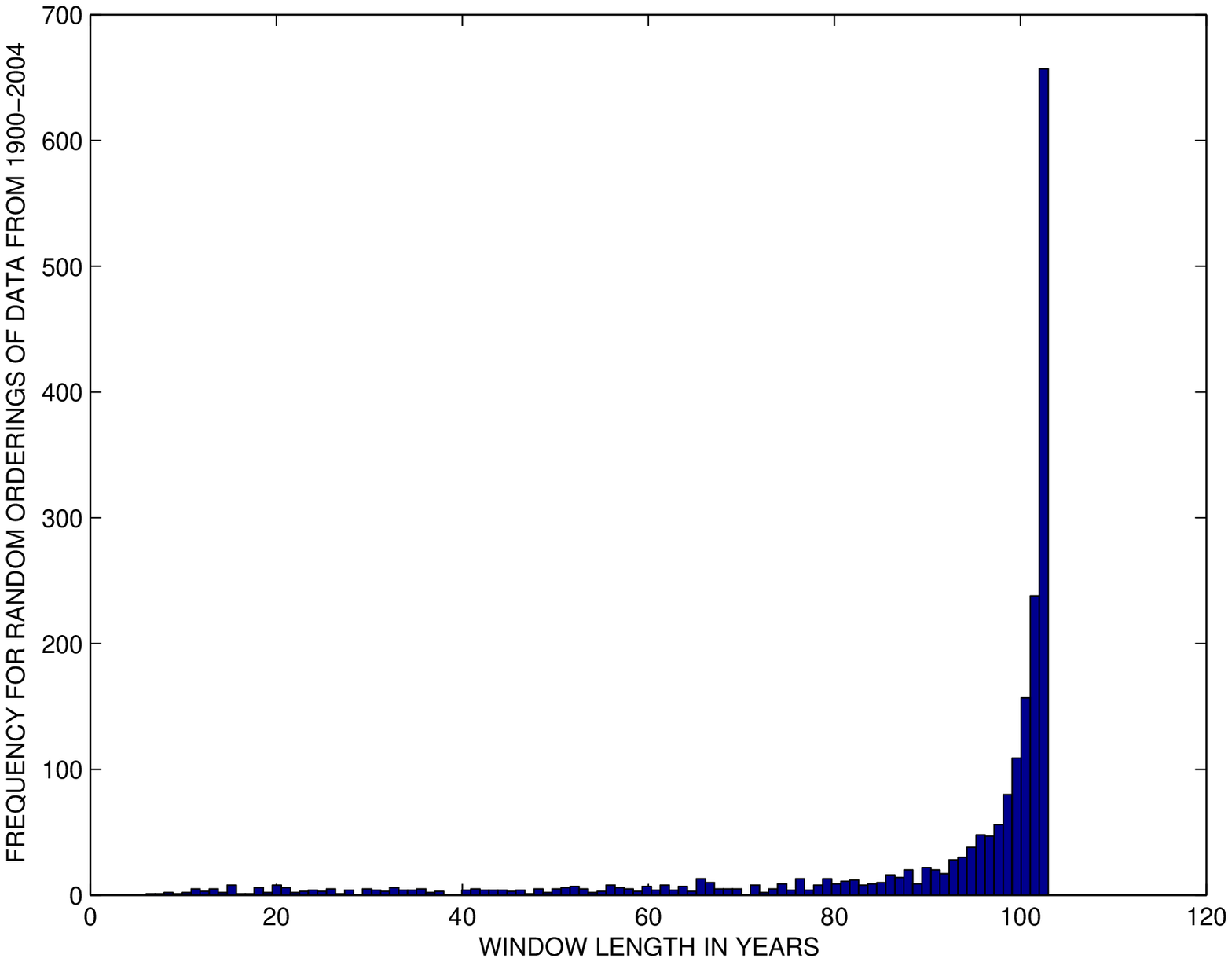}}
  \end{center}
  \caption{
As for figure~\ref{f3}, but for data from 1900 to 2004.
}
  \label{f5}
\end{figure}

\newpage

\end{document}